%% file: main.tex
\documentclass[letterpaper]{article} 
\usepackage{aaai2026}  
\usepackage{times}  
\usepackage{helvet}  
\usepackage{courier}  
\usepackage[hyphens]{url}  
\usepackage{graphicx} 
\usepackage{natbib}  
\usepackage{caption} 
\usepackage{algorithm}
\usepackage{algorithmic}
\usepackage{xcolor}
\usepackage{rotating}
\usepackage{tabularray}
\usepackage{subcaption}
\usepackage{dirtytalk}
\usepackage{amsmath}
\usepackage{booktabs}

\usepackage{newfloat}
\usepackage{listings}
\DeclareCaptionStyle{ruled}{labelfont=normalfont,labelsep=colon,strut=off} 
\floatstyle{ruled}
\newfloat{listing}{tb}{lst}{}
\floatname{listing}{Listing}

\title{Examining the Effect of Explanations of AI Privacy Redaction in AI-mediated Interactions}
\author{
    Roshni Kaushik\textsuperscript{\rm 1},
    Maarten Sap\textsuperscript{\rm 2},
    Koichi Onoue\textsuperscript{\rm 1}
}
\affiliations {
    \textsuperscript{\rm 1}Fujitsu Research of America,
    \textsuperscript{\rm 2}Carnegie Mellon University\\
    rkaushik@fujitsu.com, msap2@andrew.cmu.edu, konoue@fujitsu.com\\
}

\begin{document}

\maketitle

\begin{abstract}
\input{text/00_abstract}
\end{abstract}

\definecolor{ResearcherColor}{HTML}{A02B93}
\definecolor{CoordinatorColor}{HTML}{156082}
\definecolor{CollaboratorColor}{HTML}{196B24}
\definecolor{RevisionColor}{HTML}{000000}

\section{Introduction}
\label{sec:intro}
\input{text/01_intro}

\section{Related Work}
\label{sec:related_work}
\input{text/02_related_work}

\section{AI-Redaction System Design}
\label{sec:system_design}
\input{text/03_system_design}

\section{User Study Design}
\label{sec:study_design}
\input{text/04_study_design}

\section{Results and Discussion}
\label{sec:results}
\input{text/05_results}

\section{Conclusion}
\label{sec:conclusion}
\input{text/06_conclusion}

\section*{Ethical Considerations}
Our user study protocol was reviewed and approved by an internal ethics committee. All online study participants were compensated for their time, and their Prolific IDs were collected solely for compensation purposes.

\appendix
\include{text/07_appendix}

\bibliography{bibliography}

\end{document}

%% file: text/00_abstract.tex
AI-mediated communication is increasingly being utilized to help facilitate interactions; however, in privacy sensitive domains, an AI mediator has the additional challenge of considering how to preserve privacy. In these contexts, a mediator may redact or withhold information, raising questions about how users perceive these interventions and whether explanations of system behavior can improve trust. In this work, we investigate how explanations of redaction operations can affect user trust in AI-mediated communication. We devise a scenario where a validated system \emph{removes} sensitive content from messages and \emph{generates explanations} of varying detail to communicate its decisions to recipients. We then conduct a user study with $180$ participants that studies how user trust and preferences vary for cases with different amounts of redacted content and different levels of explanation detail. Our results show that participants believed our system was more effective at preserving privacy when explanations were provided ($p<0.05, \text{Cohen's } d \approx 0.3$). We also found that contextual factors had an impact; participants relied more on explanations and found them more helpful when the system performed extensive redactions ($p<0.05$, Cohen's $f \approx 0.2$). We also found that explanation preferences depended on individual differences as well, and factors such as age and baseline familiarity with AI affected user trust in our system. These findings highlight the importance and challenge of balancing transparency and privacy in AI-mediated communications and suggest that adaptive, context-aware explanations are essential for designing privacy-aware, trustworthy AI systems.

%% file: text/01_intro.tex
As AI systems are increasingly integrated into everyday communication, researchers have been exploring their role not only as assistants that respond to direct user queries but also as intermediaries that shape the flow of information between people. For example, AI systems are already being used to support multilingual communication through real-time translation \cite{shahmerdanova_artificial_2025}, to simplify specialized texts for broader audiences \cite{araujo_simplifying_2023}, and to remove sensitive information from shared records in healthcare \cite{caine_patients_2013}. In these cases, the AI is no longer a passive tool, but is an active mediator making decisions about what to share with whom and how to present those decisions. Increasing efforts to develop AI intermediaries suggest their potential to become an important participant in human-human interactions.

One particularly challenging and underexplored setting for AI-mediated communication is the preservation of privacy. In high-stakes domains, different parties may have different permissions to access sensitive data. For example, a clinician might share records with a specialist, while an AI system redacts personally identifiable information to comply with privacy regulations. In a workplace setting, an AI could help managers and employees collaborate while removing sensitive HR details. In research collaborations, an AI might enable data sharing between institutions by removing data that particular parties should not have access to. In these settings, the AI mediator does not simply summarize or transmit information, but actively redacts content. This creates a fundamentally different interaction paradigm where the AI alters messages while attempting to preserve trust and understanding between the humans involved. Unlike traditional settings where AI outputs are evaluated for accuracy, the system must balance both privacy protection and interaction satisfaction, raising questions about how users interpret and respond to AI-mediated redactions.

\textbf{In this work, we examine how humans perceive AI interventions in privacy-preserving scenarios, focusing on AI explanations of redactions, which has been underexplored in previous work.} This setting provides a new lens on trust in AI systems, and users must reason about what the AI shows and withholds. In contrast to standard explanation settings, where explanations justify observed model outputs, explanations in these scenarios must justify intentional transformations of the content. This shifts the user's task from evaluating the correctness of visible outputs to reasoning about potentially missing information and correctness of the system's redaction, which introduces additional challenges for calibrating system trust and reliance.

We focus on explanations as they are increasingly incorporated into AI outputs (e.g. reasoning models such as deepseek-r1 \cite{guo_deepseek-r1_2025}, openai-o1 \cite{jaech_openai_2024}) and serve as the primary mechanism for explaining the AI's intervention. Prior work suggests that explanations can act as double-edged swords; they may increase perceived transparency and trust, but they can also lead to over-reliance or reduce user scrutiny of the system \cite{park_critical_2025}. This tension is particularly powerful in privacy-preserving scenarios, where users must decide whether to accept or question an AI's redaction decisions.

While explanations can clarify why information was withheld and help users understand the system's behavior, they also introduce design trade-offs. Explanations that are absent or too general may make the AI's actions seem arbitrary or confusing, while explanations that are too detailed risk overwhelming users or inadvertently disclosing private information. This tension is particularly seen in privacy-sensitive contexts, where longer outputs increase the probability of exposing memorized or private content \cite{carlini_secret_2019}, highlighting the conflict between transparency and privacy.

\begin{figure}[t]
    \centering
    \includegraphics[width=\columnwidth]{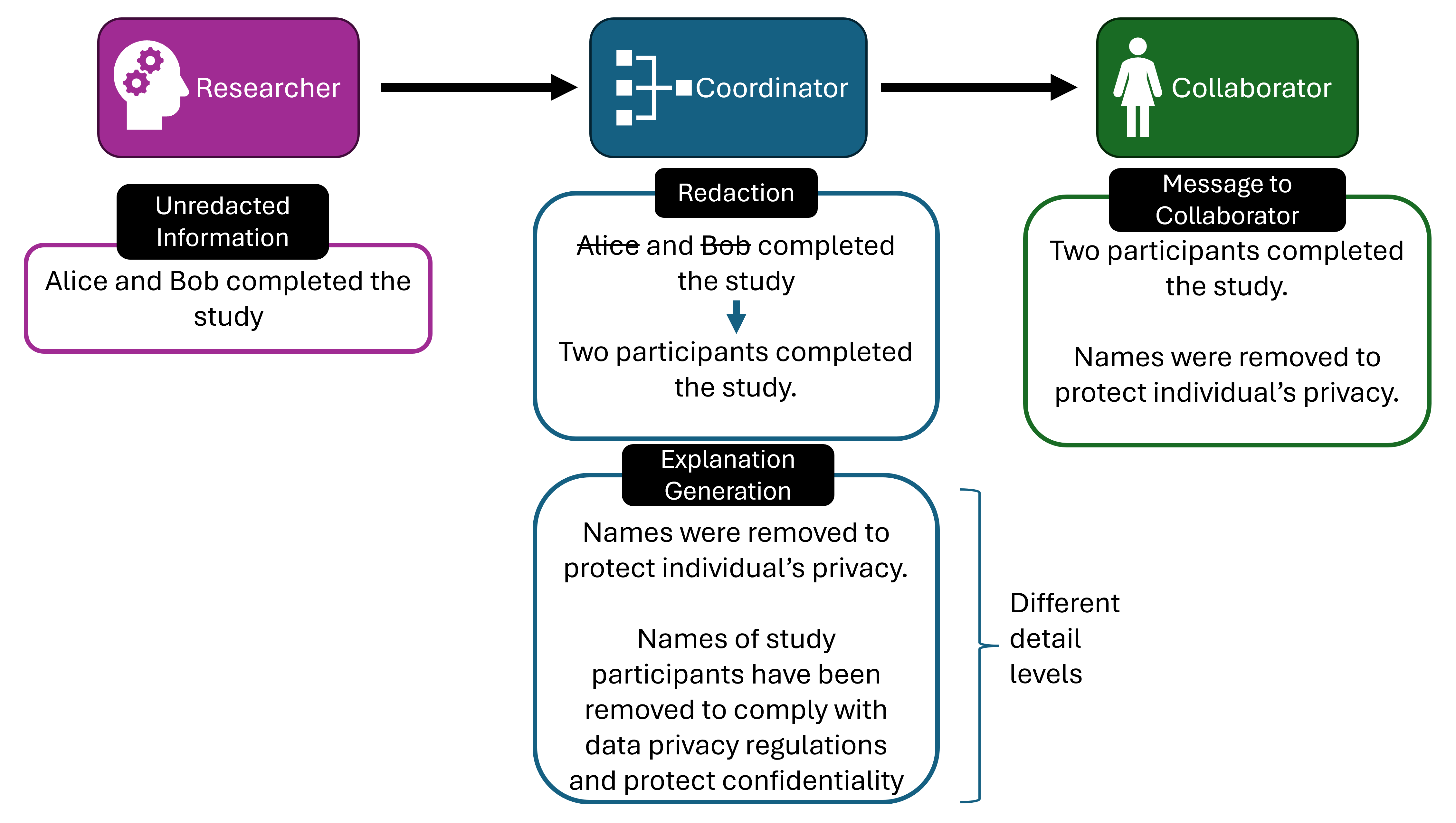}
    \caption{Conceptual overview of AI-mediated communication. The AI \textcolor{CoordinatorColor}{\textbf{coordinator}} mediates between a \textcolor{ResearcherColor}{\textbf{researcher}} and \textcolor{CollaboratorColor}{\textbf{collaborator}} by redacting sensitive information and providing an explanation to the recipient.}
    \label{fig:intro_figure}
\end{figure}

To study this, we craft a scenario where an AI mediator is tasked with removing any sensitive information that is sent by one party with information access before sending the redacted version to another party without information access. As previous work has shown that information asymmetry can negatively impact the trust of those with access to less information \cite{bertolin_information_2008}, we designed the AI mediator to provide an explanation to the party without access to improve their trust and explain the redaction actions that the system performed.

More concretely, we devised a communication setup between two parties: (1) an external \textcolor{CollaboratorColor}{\textbf{collaborator}}, a human participant without access to private information, and (2) a \textcolor{ResearcherColor}{\textbf{researcher}}, simulated for this work and with access. An AI mediator (\textcolor{CoordinatorColor}{\textbf{coordinator}}) facilitates communication between the two parties. As shown in the example in 
Figure \ref{fig:intro_figure}, the \textcolor{CoordinatorColor}{\textbf{coordinator}} redacts private information from the \textcolor{ResearcherColor}{\textbf{researcher}}'s messages and provides explanations of its decisions to the \textcolor{CollaboratorColor}{\textbf{collaborator}}. We verified the technical robustness of our AI mediator by validating its privacy preservation capabilities in our setup. 

Using our verified system, we conducted a study with $180$ participants to explore how user trust, system understanding, and explanation preferences change by systematically varying both the amount of sensitive information required to be redacted and the level of explanation detail. Specifically, we ask three research questions:
\begin{itemize}
    \item Does providing an explanation to the \textcolor{CollaboratorColor}{\textbf{collaborators}} improve their trust in and understanding of the system?
    \item How does the amount of sensitive information removed from the \textcolor{ResearcherColor}{\textbf{researcher}}'s message affect the \textcolor{CollaboratorColor}{\textbf{collaborators}}' explanation preferences?
    \item Do individual differences (e.g., demographic or attitudinal factors) influence trust and explanation preferences?
\end{itemize}

Our results show that participants believed our system was more effective at preserving privacy when explanations were provided ($p<0.05, \text{Cohen's } d \approx 0.3$). We also found that contextual factors had an impact; participants relied more on explanations and found them more helpful when the system performed extensive redactions ($p<0.05, \text{Cohen's } f \approx 0.2$). Individual characteristics further influenced trust; for example, those with a baseline familiarity or trust in AI/LLM systems were also more likely to trust our system.

Explanations for AI-mediated communication that preserve privacy present a complex challenge, requiring a careful balance between transparency and protection of sensitive information. Our work highlights this difficulty by exploring how different redaction strategies and explanation styles affect user understanding and trust. By examining these interactions, we shed light on design principles and open questions that can guide future research on creating AI-mediated systems that are both privacy-aware and user-friendly.

%% file: text/02_related_work.tex
Previous research relevant to our work includes contributions from AI-mediated interaction, privacy-preserving communication, explanation design, and trust in AI systems. Together, these areas highlight interesting intersections between protecting sensitive content and maintaining user understanding.

\subsection{AI Mediation of Human Communication}
Research in HCI has begun to characterize AI not just as a tool but as an active intermediary that modifies and reshapes interpersonal exchanges. Hancock et al. offer a foundational framework for AI-mediated communication, where intelligent agents modify or generate messages on behalf of a human \cite{hancock_ai-mediated_2020}. They articulate key characteristics, such as the magnitude of involvement, media type (text, audio, video, etc.), optimization goal, autonomy, and role orientation (who the AI is operating on behalf of). This allows us to understand the various ways that AI can intervene in human communication. Other research has shown that trust perceptions in senders diminish when recipients know that AI is involved, but surprisingly trust increased when AI wrote interpersonally rather than transactionally \cite{liu_will_2022}.

Psychology studies have also revealed how humans attribute agency, responsibility, and trust in mediated contexts. Users often reassess blame when AI mediates, and trust in the human communicator remains high. However, some responsibility can shift to the AI when interactions fail, known as the moral crumple zone phenomenon \cite{hohenstein_ai_2020}. Frameworks such as AI-Mediated Exchange Theory \cite{ma_ai-mediated_2020} explain these types of trust shifts in the balance of social exchange. Research on human perceptions based on the author of certain texts found that only when participants thought they saw a mix of AI and human-written profiles, they mistrusted hosts whose profiles were labeled as or suspected to be written by AI \cite{jakesch_ai-mediated_2019}. These works position AI mediation as not only a technical innovation, but also a social intervention that has implications for trust and sense of agency. However, prior work has not yet examined how AI mediation shapes trust and understanding when the system is actively altering messages for privacy reasons. Our work addresses this gap by studying explanations as a mechanism to support trust in privacy-preserving AI mediation.

\subsection{Privacy-Preserving Communication}
Privacy is a critical concern in mediated communication, especially in domains like healthcare, workplace collaboration, and research. Nissenbaum's foundational framework for contextual integrity argues that privacy is maintained when information flow follows contextual norms \cite{nissenbaum_privacy_2004}. HCI research has explored many methodologies to protect sensitive information while enabling collaboration, including for selective redaction, access controls, and data anonymization \cite{caine_patients_2013}.

Advances in natural language processing have enabled automatic detection and redaction of private information, but these methods face trade-offs between privacy and utility \cite{sousa_how_2023, pal_empirical_2024}. Redacting too much information can reduce message clarity, but not redacting enough can lead to privacy leaks. Security and privacy research has begun to systematically evaluate these trade-offs. For example, recent work has investigated the vulnerabilities of LLMs in the extraction of personal information and the effectiveness of redaction as a countermeasure \cite{liu_evaluating_2024, cheng_effective_2025}. Studies have also explored how contextual integrity can serve as a framework for auditing privacy risks with language models \cite{mireshghallah_can_2023}. 

From an HCI perspective, there have been calls for research on how LLM design paradigms affect disclosure behaviors, mental models and preferences for privacy control, and the design of tools, systems, and artifacts that empower users to reclaim ownership over their personal data \cite{li_human-centered_2024}. Setting the correct expectations for AI can drastically change the users' perspective of the system and interaction \cite{kocielnik_will_2019}, and adding design paradigms, such as explanations, must balance informativeness while preserving privacy. Other efforts have emphasized the need for user empowerment in the privacy preservation process through interactive AI tools \cite{sun_empowering_2024}. However, it remains unclear how different explanation strategies interact with the varying levels of redaction that a system might perform to impact user trust and comprehension. Our work addresses this open question by systematically studying these factors in an AI-mediated, privacy-preserving communication scenario.

\subsection{Explanations in AI-mediation}
Explanations are a central mechanism to help users understand AI behavior. Work in explainable AI has shown that explanations influence how people understand, predict, and evaluate system decisions \cite{miller_explanation_2019}. HCI researchers emphasize that explanations must align with users' goals and preferences to improve human-AI collaboration \cite{ehsan_automated_2019}. Design-oriented reviews of human–XAI interaction further highlight that explanations should be interactive, context-aware, and tailored to the needs of different users to support effective decision-making \cite{chromik_human-xai_2021}. When AI acts as a mediator, explanations can clarify why information was altered or withheld and can help users maintain a coherent understanding of the interaction. Recent work has also found that AI-generated explanations could have a large impact on human trust, and factors such as the AI background of the user can heavily influence the interpretation of explanations \cite{ehsan_who_2024}. Prior research also suggests that individual user characteristics can meaningfully shape how people interpret and rely on explanations in decision-making contexts \cite{khadar_explain_2025}. This motivates our addition of background and personality questions as part of our user questionnaire.

Beyond usability concerns, explanations are also closely tied to broader discussions of accountability and ethics in AI systems. Researchers argue that transparency mechanisms such as explanations are essential for holding algorithmic systems accountable and enabling meaningful oversight \cite{busuioc_accountable_2021}. At the same time, ethical analyses of AI systems emphasize the importance of transparency alongside bias mitigation and responsible system design to ensure trustworthy deployment of AI technologies \cite{mensah_artificial_2023}.

However, explanations in mediated communication can raise unique challenges. Explanations can act as double-edged swords by increasing perceived transparency and trust, but also promoting to over-reliance or reduce user scrutiny of the system's outputs \cite{park_critical_2025}. This tension is particularly powerful in privacy-preserving scenarios, where users must decide whether to accept or question an AI's decision to redact information. Surveys of privacy-preserving explainable AI highlight that explanations themselves may leak sensitive details if not designed carefully \cite{dosilovic_explainable_2018}. Studies on attacks against explanations from models have shown that adversaries can sometimes infer private information from outputs that appear to be anonymized \cite{shokri_privacy_2021}. Reducing the amount of information disclosed to combat this can come at the cost of informativeness \cite{nguyen_survey_2024}. Systems have been developed that help users identify sensitive information and summarize relevant privacy policies, which has improved users’ understanding of data
practices and privacy risks \cite{chen_clear_2025}. Additionally, prior work \cite{krsek_measuring_2025} has developed a tool to help users catch accidental private information leakage and found that the AI should account for context and user differences when providing explanations for privacy modifications.

Together, these findings suggest that explanations can have a powerful impact in privacy contexts, but issues such as context-dependence and accidental privacy leakage must also be considered. How people actually experience and evaluate explanations when an AI mediates their communication with another human is an underexplored area. Our work addresses this gap by examining how both explanation detail and redaction amount shape preferences, trust, and understanding in a privacy-sensitive, AI-mediated setting.

\subsection{Trust in AI and Human-AI Collaboration}
Trust has long been recognized as a critical factor in the way people interact with intelligent systems. Lee et al.~\cite{lee_trust_1992} explored how trust influences control strategies and how responsibility was allocated between humans and systems. Later frameworks such as the Intelligent Systems Technology Acceptance Model integrated transparency and acceptance into intelligent system models \cite{vorm_integrating_2022}. In human-AI collaboration, maintaining the appropriate level of trust is critical. Over trust can lead to overreliance, while under trust can lead to misunderstandings and lower use of the system.

Recent HCI research explores these dynamics in AI-mediated decision making. Trust must be understood from the perspective of those who use AI systems and from those affected by their outputs, resulting in a complex multi-stakeholder dynamic \cite{vereschak_trust_2024}. A systemic overview of models and trust measures in \cite{ueno_trust_2022} details the variety of approaches that can be applied in different scenarios. Explainability is often explored as a way to calibrate user trust and can help users understand system outputs and limitations \cite{kim_help_2023}. Prior work also distinguishes between attitudinal trust (a user's belief that a system is reliable) and behavioral reliance (the extent to which users actually follow or act on the system's outputs). Recent research on explainable reliance-aware user interfaces emphasizes designing explanations that help users decide when to accept, question, or overrule AI recommendations \cite{de_souza_filho_where_2024}. We focus on attitudinal trust in this work.

However, most of this research considers trust in systems that provide information rather than those who also can choose to withhold it. When an AI redacts information for privacy reasons, users must assess the reliability of the remaining information under conditions of incomplete information, which then shapes their trust in the system. Our work addresses this open research question by investigating trust in AI mediation, where explanations and privacy-preserving redactions intersect.

%% file: text/03_system_design.tex
\begin{figure}[ht]
    \centering
    \includegraphics[width=\columnwidth]{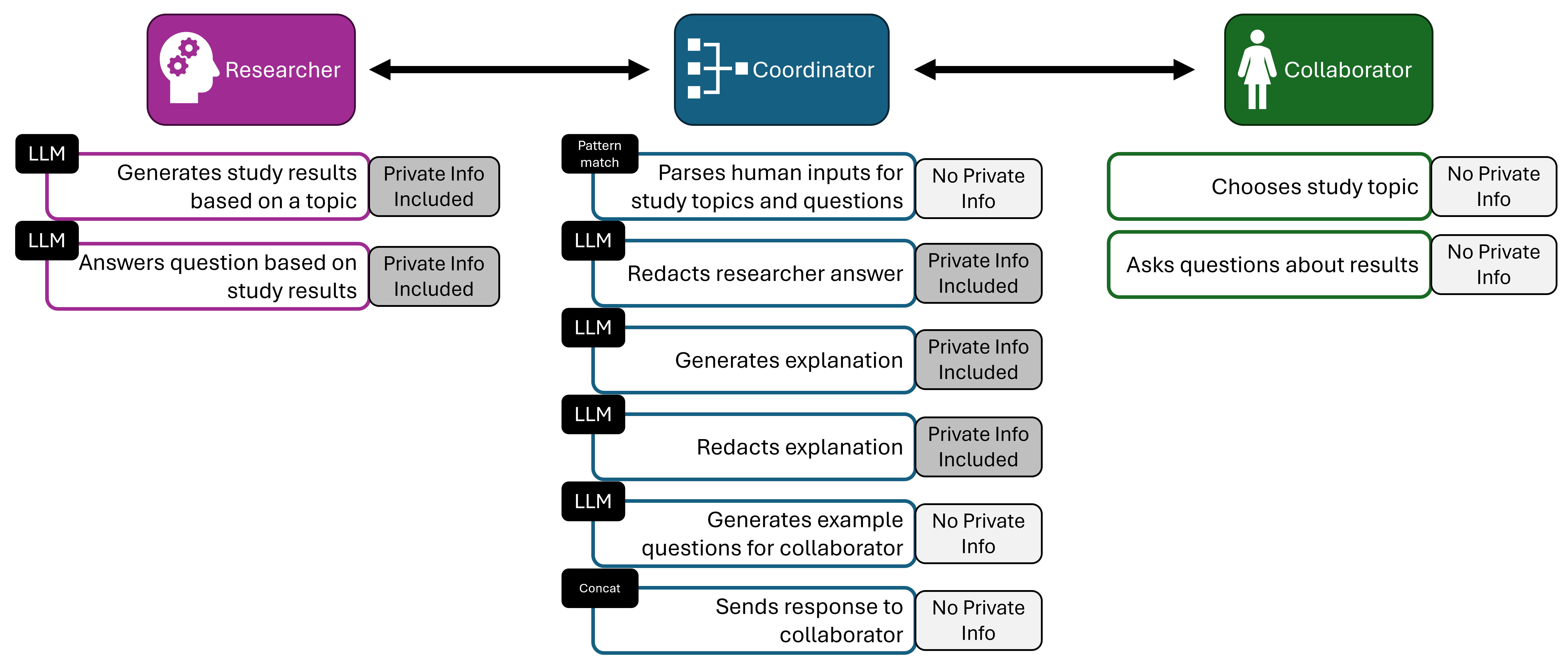}
    \caption{Diagram of the interactions between the \textcolor{ResearcherColor}{\textbf{researcher}}, \textcolor{CollaboratorColor}{\textbf{collaborator}}, and \textcolor{CoordinatorColor}{\textbf{coordinator}}, with tasks performed by each that include and do not include private information indicated.}
    \label{fig:system_diagram}
\end{figure}

We devise a scenario where our system (Figure \ref{fig:system_diagram}) enables human-AI collaboration and mediates between two human roles: \textcolor{ResearcherColor}{\textbf{researcher}} and external \textcolor{CollaboratorColor}{\textbf{collaborator}}. The \textcolor{ResearcherColor}{\textbf{researcher}} has performed some research (such as a user study) that they would like to share with the \textcolor{CollaboratorColor}{\textbf{collaborator}}. However, there is information in the results, such as names and genders, that the \textcolor{CollaboratorColor}{\textbf{collaborator}} cannot have access to. Our system sits between these two roles and performs privacy-preserving operations on the information the \textcolor{ResearcherColor}{\textbf{researcher}} is sharing to ensure that any information the \textcolor{CollaboratorColor}{\textbf{collaborator}} does not have access to is removed. The \textcolor{CollaboratorColor}{\textbf{collaborator}} can ask questions about the information to the \textcolor{ResearcherColor}{\textbf{researcher}}, and our system will remove any private information in the \textcolor{ResearcherColor}{\textbf{researcher}}'s answer before sending it to the \textcolor{CollaboratorColor}{\textbf{collaborator}}.

In the system we implemented for this work, the human is the external \textcolor{CollaboratorColor}{\textbf{collaborator}}, and the \textcolor{ResearcherColor}{\textbf{researcher}} is simulated by an LLM, as we are interested in the \textcolor{CollaboratorColor}{\textbf{collaborator}}'s understanding, trust, and explanation preferences. We focus on the \textcolor{CollaboratorColor}{\textbf{collaborator}} perspective because \textcolor{CollaboratorColor}{\textbf{collaborator}}s, unlike \textcolor{ResearcherColor}{\textbf{researcher}}s, must rely on AI explanations to interpret missing information, making their explanation needs particularly important in privacy-sensitive interactions \cite{miller_explanation_2019, hoff_trust_2015}. Additionally, we consider a relatively simple redaction task, the removal of Personally Identifiable Information (PII); however, our approach could be extended in future work to more complex privacy-preserving scenarios.

Figure \ref{fig:system_diagram} illustrates the tasks that the \textcolor{ResearcherColor}{\textbf{researcher}}, \textcolor{CollaboratorColor}{\textbf{collaborator}}, and \textcolor{CoordinatorColor}{\textbf{coordinator}} perform in this interaction, as well as which of these tasks include and do not include private information. The \textcolor{ResearcherColor}{\textbf{researcher}} and the \textcolor{CoordinatorColor}{\textbf{coordinator}} both have access to the private information. The \textcolor{ResearcherColor}{\textbf{researcher}} shares the unredacted study results and answers to the \textcolor{CollaboratorColor}{\textbf{collaborator}} questions. The \textcolor{CoordinatorColor}{\textbf{coordinator}} takes that information, removes any private information, and generates an explanation for the \textcolor{CollaboratorColor}{\textbf{collaborator}} that explains the privacy-preserving operations that were carried out on the \textcolor{ResearcherColor}{\textbf{researcher}}'s information. The \textcolor{CoordinatorColor}{\textbf{coordinator}} also performs some tasks for the purposes of our particular interaction and study that do not include private information, such as parsing the \textcolor{CollaboratorColor}{\textbf{collaborator}}'s input, compiling the redacted response for the \textcolor{CollaboratorColor}{\textbf{collaborator}}, and generating some sample questions that could be interesting to ask the \textcolor{ResearcherColor}{\textbf{researcher}}. Figure \ref{fig:chat_outline} (see Section \ref{sec:appendix-example_interaction}) includes an excerpt from an example interaction between the \textcolor{CoordinatorColor}{\textbf{coordinator}} and the \textcolor{CollaboratorColor}{\textbf{collaborator}}. GPT 4.1 was used for all LLM calls in this work, and more prompting details are included in Section \ref{sec:appendix-prompting_details}.

\subsection{Study Topics and Original Information}
We generated five different study topics that did not require specialized knowledge to comprehend and could be interesting/applicable to most people. These topics were (1) the impact of gardening on mental health, (2) the effects of exercise on sleep, (3) the relationship between diet and mood, (4) the influence of social media on self-esteem, and (5) the role of mindfulness in stress reduction.

For each study topic, we generate a set of \textcolor{RevisionColor}{mock but realistic} user study results that included some private information about participants (e.g., names, ages, genders, ethnicities) and some general information that was not private (e.g., amount of sleep each night). We defined private information using standard NLP/HCI criteria\cite{schwartz_pii_2011, song_understanding_2025}, treating any attribute that could directly or indirectly identify a participant as private, and did not apply formal guarantees like k-anonymity. \textcolor{RevisionColor}{Both the study topic and study results were with GPT 4.1, and we manually verified that the generated information included both private and general information.}

\subsection{Questions and Amount of Sensitive Information}
We then generated $25$ questions in three different redaction categories: high, moderate, and low. These redaction categories represent the amount of sensitive information present in the \textcolor{ResearcherColor}{\textbf{researcher}}'s answer to a question that needs to be removed before it can be sent to the \textcolor{CollaboratorColor}{\textbf{collaborator}}. 
Three examples of questions the \textcolor{CollaboratorColor}{\textbf{collaborator}} are included below, one from each redaction category.
\begin{itemize}
    \item \textbf{High}: What did the oldest participant say about how exercise affected their sleep?
    \item \textbf{Moderate}: How did women in the study describe their gardening experiences compared to indoor activities?
    \item \textbf{Low}: How did participants describe the difference in their sleep quality between the two conditions?
\end{itemize}
The answer to the high redaction question results in identifying this participant, so most/all of the answer should be removed. In contrast, the answer to the low redaction question requires little to no removal of sensitive information. \textcolor{RevisionColor}{All questions and answers were generated with GPT 4.1, and we manually verified that the answers to each question were correct based on the corresponding user study results. Additionally, we ensured that the questions were realistic and representative of plausible \textcolor{CollaboratorColor}{\textbf{collaborator}} queries by verifying they were grounded in the types of analyses typically performed on user study data (e.g., comparing participant groups, examining condition differences, summarizing qualitative responses) \cite{braun_using_2006}.}

\textcolor{RevisionColor}{We next created an LLM judge to validate that the amount of sensitive information present in the answers to these questions was sufficiently different in each redaction category. Note that this is a different LLM judge to the one presented below to verify privacy protection. We found that the questions in each redaction category were sufficiently different; specifically the generated questions in the high redaction category contained more sensitive information than the moderate category and the moderate category contained more sensitive information than the low category. Further details are included in Section \ref{sec:appendix-redaction_categories}.}

\subsection{Redacting Sensitive Information from Answers}
\textcolor{RevisionColor}{We then redacted sensitive information from all the \textcolor{ResearcherColor}{\textbf{researcher}} answers using GPT4.1. We used a validated LLM judge (see Section \ref{sec:systemdesign-data_validation}) to test whether there was any private information remaining in the full set of redacted answers. We found that our system protected private information for at least $98\%$ of the data in each redaction category (more details in Section \ref{sec:appendix-redacted_answer_privacy}).}

\subsection{Data Validation of Privacy Preservation}
\label{sec:systemdesign-data_validation}
\textcolor{RevisionColor}{To begin the validation process, we first assigned two human coders to determine whether privacy was preserved for a subset of data ($60$ randomly selected pieces of data from each redaction category). The human coders agreed for all of the data they evaluated, which allows us to rely on their responses for this subset.}

\textcolor{RevisionColor}{We next set up an LLM judge to test whether private information was present after redaction was performed, motivated by prior research that shows that LLMs are effective at identifying and handling PII \cite{krco_rat-bench_2026}. We tested two LLM judges (GPT 4.1 and Claude Opus 4.7) and compared their evaluations to each other on the same data subset and found perfect agreement between judges ($100\%$), indicating a lack of self-bias, which could occur when generating and evaluating with the same model. Finally, we verified alignment between the LLMs and human coders and found high agreement between them (specifically $93.9\%$ agreement between the human coders and GPT 4.1). This indicates that the result of our LLM judge is reliable in addition to not being a result of self-bias.}

\subsection{Explanations of Different Types}
Once we generated the answer and redacted answer to each question, we generated explanations in two distinct styles. Building on prior work \cite{larasati_effect_2020} which explored how explanation styles influence trust, we further elaborate the distinction between \emph{general} and \emph{thorough} explanations as two points along a broader explanation design spectrum. In particular, we distinguish these styles by length and their level of abstraction and completeness.

The \emph{general} explanation is concise (1-2 sentences in our setting) and communicates high-level information about why information was withheld, without detailing the specific steps the system performed. In contrast, the \emph{thorough} explanation is longer (3-5 sentences) and provides a more complete, step-by-step account of how the system identified and removed private information. This distinction allows us to compare explanations that prioritize conciseness against those that prioritize transparency and detail. Examples of these two explanation styles for the same answer and redacted answer are shown in Section \ref{sec:appendix-explanation_examples}.

We focus on these two explanation styles to explicitly explore the trade-offs introduced by increasing explanation detail. We hypothesize that more thorough explanations may reduce user misunderstanding and increase perceived transparency. However, longer and more detailed explanations increase the likelihood that an LLM may inadvertently reveal sensitive information or overwhelm the user with too much detail. In contrast, more general explanations may better preserve privacy and reduce user cognitive load, but risk not including enough detail.

As we show in Section \ref{sec:appendix-redacted_explanation_privacy}, private information is being leaked in the thorough explanations, but not in the general explanations, with the validation performed by the verified LLM judge in Section \ref{sec:systemdesign-data_validation}. The longer explanations led to a greater amount information leakage. We therefore decided to redact private information from the generated explanations using the same GPT4.1 prompt used to redact the answers. After information redaction, we found that there was no private information remaining in the explanations (see graph in Section \ref{sec:appendix-redacted_explanation_privacy}).

%% file: text/04_study_design.tex
To explore our research questions, we conducted a user study (protocol approved by an internal ethics committee). The participants interacted with an LLM-simulated \textcolor{ResearcherColor}{\textbf{researcher}} through the AI \textcolor{CoordinatorColor}{\textbf{coordinator}}, which applied appropriate redactions and provided explanations. Our two independent variables were explanation style (no explanation, general, thorough) and redaction needed (high, moderate, low). Through a survey, we measured our dependent variables of perceived trust in the system, subjective understanding of the system, and explanation preferences. 

\subsection{Participants}
We first performed a power analysis to determine the appropriate number of participants to recruit. With three explanation styles, we determined that $180$ participants ($60$ per group) should allow us to detect between a small effect ($f=0.20$ results in a power of $0.661$) and medium effect ($f=0.5$ results in a power of $1.000$). We recruited all our participants from Prolific with a qualification of being fluent in English and compensation of $\$4$ per participant. Their average age was $33.0$, with $1.1\%$ aged $18-19$, $45.7\%$ aged $20-29$, $30.9\%$ aged $30-39$, $13.1\%$ aged $40-49$, $6.3\%$ aged $50-59$, and $2.3\%$ aged $60-69$. $52.6\%$ identified as male and $47.4\%$ as female. $3.4\%$ identified as Asian, $70.9\%$ as Black, $18.3\%$ as White, and $7.4\%$ as Mixed.

\subsection{Study Procedure}
The participants first completed a brief introduction process that included instructions explaining their role as the \textcolor{CollaboratorColor}{\textbf{collaborator}} interacting with a \textcolor{ResearcherColor}{\textbf{researcher}} through an AI \textcolor{CoordinatorColor}{\textbf{coordinator}}. They were informed that some information might be withheld or modified from the \textcolor{ResearcherColor}{\textbf{researcher}}'s responses and that the \textcolor{CoordinatorColor}{\textbf{coordinator}} might provide some explanation for those modifications. They then completed a 5-question multiple choice quiz to ensure they understood the scenario, where they had to complete each question correctly (with multiple attempts allowed).

The participants were randomly sorted into one of three explanation conditions: no explanation, general explanation, and thorough explanation. The \textcolor{CoordinatorColor}{\textbf{coordinator}} then presented them with a list of topics and instructed them to choose one they found most interesting (top of Figure \ref{fig:chat_outline}). The \textcolor{CoordinatorColor}{\textbf{coordinator}} then responded with a summary of the findings on that topic and a list of questions that the participant could choose from to follow up. These questions were from one of the redaction categories (high, moderate, or low). The participant then chose a question from the provided list, and the \textcolor{CoordinatorColor}{\textbf{coordinator}} responded with the redacted \textcolor{ResearcherColor}{\textbf{researcher}} answer, a redacted explanation based on the participant's explanation condition, and a set of new follow-up questions in a different redaction category. This process continued for three questions, and the order of the redaction categories was randomized for each participant. This resulted in a between-subjects design for the first independent variable of explanation detail and a within-subjects design for the second independent variable of redaction amount.

\subsection{Measures and Hypotheses}

\begin{figure}[t]
    \centering
    \includegraphics[width=\columnwidth]{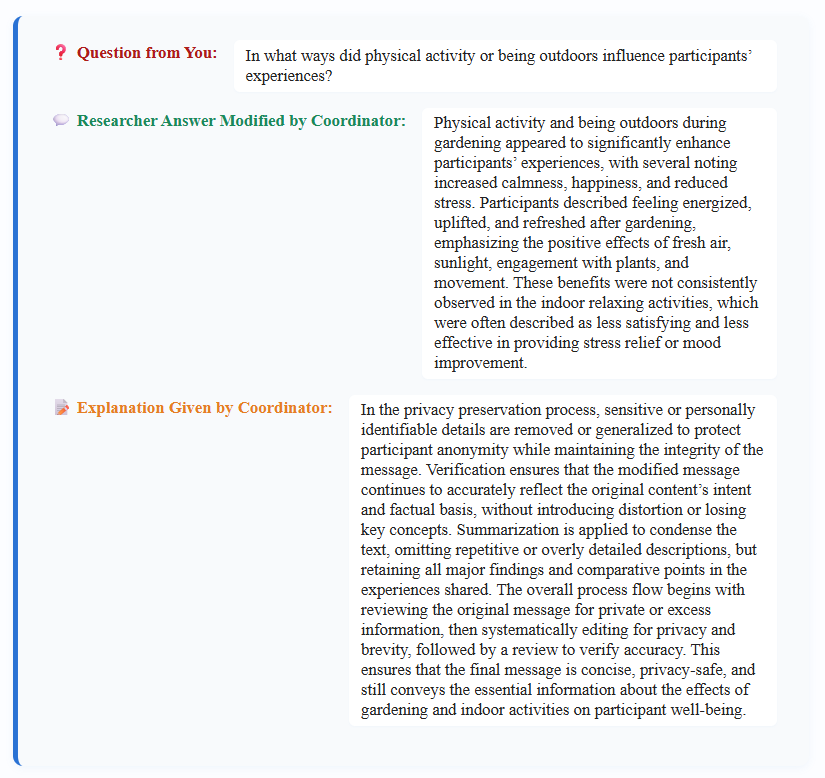}
    \caption{Screenshot of the question, redacted answer, and explanation (if provided) that are included for each per-interaction evaluation group}
    \label{fig:per_interaction_survey}
\end{figure}

After three questions and answers, the participant was redirected to the survey, which had three sections. For the first two sections, a portion of the evaluation items were modified from \cite{hoffman_metrics_2019} and \cite{jian_foundations_2000}, including an explanation satisfaction scale, a trust scale for the explainable AI context, and a checklist for trust between people and automation. The full questionnaires are included in Section \ref{sec:appendix-questionnaire}.

\subsubsection{Evaluation intended to explore explanation detail preferences}
\label{subsec:survey1}
The first section was a series of evaluation items per interaction, one for each of the question-answer-explanation interactions the participant experienced. Each question, modified researcher answer, and explanation (if any) given were shown to the participant (example in Figure \ref{fig:per_interaction_survey}) to refresh their memory, and we asked five items (four 5-pt Likert, one free form) for each of these interactions, resulting in a total of 15 items in this section of the survey. These items were intended to explore how the amount of redaction in an interaction affected the participant's trust and understanding in the system and explanation preferences.

\subsubsection{Evaluation intended to explore explanation detail and redaction amount interaction}
\label{subsec:survey2}
In the second section, we asked participants $8$ evaluation items ($7$ 5-pt Likert, $1$ free form) about their overall experiences with the system to gauge their general explanation preferences and how the explanations that they received affected their trust and system understanding.

\subsubsection{Evaluation intended to explore interaction between user characteristics and trust in our system}
\label{subsec:survey3}
In the last section, we included demographic evaluation items and some evaluation items about their attitudes and experience with AI. Previous research has found that background can impact trust in automated systems and the effect of explanations on that trust \cite{ehsan_who_2024}, so we wanted to obtain background information on our participants to see if there were correlations between their responses to these evaluation items and their experiences in the study.

\subsubsection{Hypotheses}
We tested three hypotheses through our user study that address our research questions.

\begin{itemize}
    \item (H1): People will prefer an explanation over no explanation
    \item (H2): Explanation preferences will vary between different redaction categories
    \item (H3): People will have different levels of trust in the system, based on their background and experiences
\end{itemize}

%% file: text/05_results.tex
\subsection{Explanations vs. No Explanations}
\begin{figure}[ht]
    \centering
    \includegraphics[width=\columnwidth]{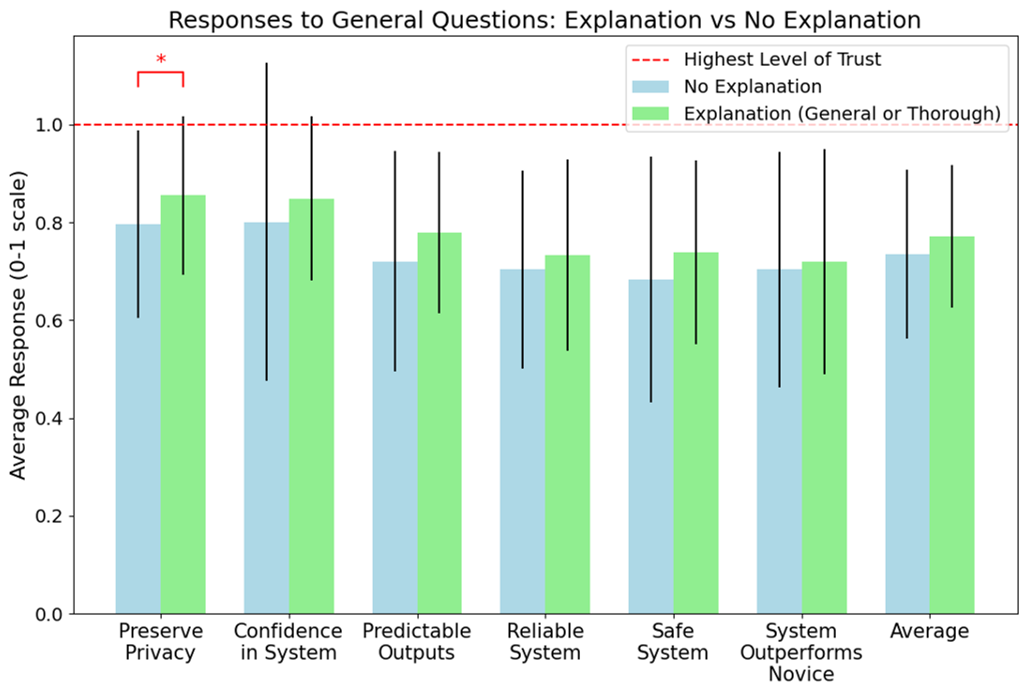}
    \caption{Responses of participants to Survey Section~\ref{subsec:survey2} about their overall experiences with the system, normalized to $0-1$. The last set of bars illustrate the average across all questions. The value corresponding to the most positive response is shown in red. Significant differences are shown in red with one asterisk for $p<0.05$.}
    \label{fig:h1_plot}
\end{figure}

We first analyzed responses from the second part of the survey (Section~\ref{subsec:survey2}), where participants answered Likert questions about their overall experiences with the system. Participants were divided into two groups for this analysis: those who received an explanation and those who did not. We normalized the ($1-5$) responses to a $0-1$ scale for plotting simplicity, and mean ratings with standard deviation bars are shown in Figure \ref{fig:h1_plot}. We performed a Welch ANOVA (due to unequal variances) to compare between these two groups. We found a significant effect of explanation on participants' trust in the system's ability to preserve privacy ($F(1, 101) = 3.95, p<0.05, \text{Cohen's } d =  0.337$). Responses to the remaining questions did not reach significance, although trends were consistent with expectations. Participants who received explanations tended to report higher confidence in the system, perceived outputs as more predictable, considered the system more reliable and safe, and rated it as outperforming a novice user, although these differences were not significant.

We also examined participants' engagement from their free-text responses. We defined engagement by identifying minimal responses, defined as containing no response or keywords such as `no comment', `none', `n/a', `nothing', `no comments', and `no additional'. In addition to identifying minimal responses, we also conducted a sentiment analysis using the Natural Language Toolkit \cite{ma2018targeted}. Results, summarized in Table \ref{tab:sentiment_analysis}, show that participants who received explanations were more engaged, less likely to provided minimal responses, and more likely to write positive comments. They were also less likely to provide negative comments, indicating that explanations not only influence quantitative trust ratings but also positively affect subjective evaluations and engagement.

\begin{table}[b]
\centering
\begin{tabular}{ccc} 
\toprule
 & \textbf{No Explanation} & \begin{tabular}[c]{@{}c@{}}\textbf{General + Thorough}\\\textbf{Explanations}\end{tabular} \\
\textbf{Minimal} & 34.0\% & 24.0\% \\
\textbf{Positive} & 56.4\% & 60.2\% \\
\textbf{Neutral} & 33.3\% & 35.2\% \\
\textbf{Negative} & 10.3\% & 4.5\% \\
\bottomrule
\end{tabular}
\caption{Analysis of the minimal responses and positive/neutral/negative responses from free-response, separated by participants who received and did not receive an explanation.}
\label{tab:sentiment_analysis}
\end{table}

\subsubsection{H1: People will prefer an explanation over no explanation}

These results provide partial support for the hypothesis. Participants who received explanations rated the system as more effective in protecting privacy than those who did not, indicating that explanations can improve trust in privacy-preserving systems.

The pattern was reinforced by participants’ open-ended responses. Table~\ref{tab:sentiment_analysis} shows that explanations increased engagement, encouraged positive comments, and reduced negative feedback. Several participants who did not receive explanations expressed a desire for more information (\say{The system sometimes gave useful and detailed answers but other times withheld too much information without explanations.}), whereas others were not bothered by limited answers (\say{I had an easy experience with the system. [I]t did what it was supposed to despite not answering that one question that had to do with age.}). Participants who received explanations highlighted the value of transparency: one noted, \say{Better to honestly say that information cannot be revealed}, while another said, \say{The system appears to be trustworthy with the user's private information.} These comments suggest that explanations signal reasoning and intent, which reassure users and support trust. 

These findings are consistent with prior research in explainable AI showing that explanations increase trust by improving perceived transparency \cite{kizilcec_how_2016}. Additionally, our results suggest that even for a relatively simple privacy-preservation task, such as PII removal, providing explanations can positively influence how users interpret and interact with the system. Beyond supporting user trust, explanations also support user agency by enabling users to assess whether system decisions align with their expectations. From a socio-technical perspective, this highlights that explanations shape not only perceptions of the system, but also the broader relationship between users and AI systems. When relevant information is withheld or explanations are unclear, users' ability to critique the system's assumptions or identify potential errors is limited. Explanations could therefore be used as mechanisms for increasing trust and as tools to support user investigation, accountability, and more informed human-AI interaction.

\subsection{Preference Variances by Amount of Redaction Needed}
\begin{figure*}[t]
    \centering
    \includegraphics[width=2\columnwidth]{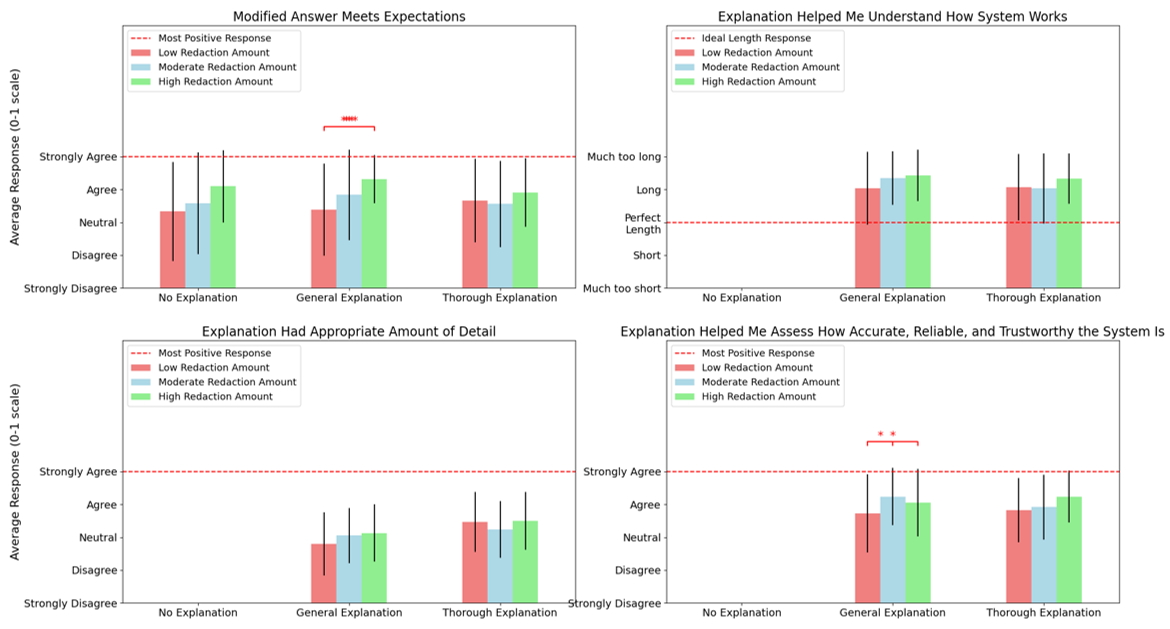}
    \caption{Responses of participants to Survey Section~\ref{subsec:survey1} about their overall experiences with the system, normalized to $0-1$. Responses are separated by the explanation received by participants and by the amount of redaction needed for the answer. The value corresponding to the most positive response is shown by a dotted line.  Significant differences are shown in red with one, two, and three asterices for $p<0.05$, $p<0.01$, and $p<0.001$, respectively.}
    \label{fig:h2_plot}
\end{figure*}

We analyze responses from the first part of the survey (Section~\ref{subsec:survey1}), where participants answered Likert questions for each of the three interactions they experienced in the study, corresponding to low, moderate, and high redaction cases. Participants were grouped according to the type of explanation they received: none, general, or thorough. Their responses ($1$–$5$) were normalized to a ($0$–$1$)  for plotting simplicity, and Figure~\ref{fig:h2_plot} shows the mean ratings with standard deviation bars for each of the four questions, illustrating how the’ perceptions changed across redaction levels.

Because the groups exhibited unequal variances, we used a Welch ANOVA with Games–Howell post hoc tests to identify significant differences, which are marked in the plot. Significant differences emerged for the first question (top-left plot), which asked how well the modified answer matched participants’ expectations. For participants who did not receive an explanation, there was a significant effect of redaction level on perceived expectation-match ($F(2, 115.63) = 5.29, p<0.01, \text{Cohen's } f =  0.218$), with high redaction cases matching expectations significantly more than low redaction cases ($p<0.01$). Participants who received a general explanation showed a similar pattern ($F(2, 105.21) = 10.35, p<0.001, \text{Cohen's } f =  0.305$), again rating high redaction cases as significantly more aligned with their expectations than low redaction cases ($p<0.001$). Across both groups, greater system intervention was associated with a stronger sense that the modified answer was appropriate or expected.

Significant differences also emerged for the fourth question (bottom-right plot), which examined how well the explanation helped participants judge the system’s accuracy, reliability, and trustworthiness. Among participants who received a general explanation, redaction level significantly affected their ability to assess the system ($F(2, 107.25) = 3.26, p<0.05, \text{Cohen's } f = 0.2035$), with moderate redaction cases yielding better assessments than low redaction cases. Participants who received a thorough explanation showed a similar effect ($F(2, 112.02) = 3.47, p<0.05, \text{Cohen's } f = 0.189$), with high redaction cases producing better assessments than low redaction ones. In both explanation conditions, the explanation was more helpful when the system performed more extensive modifications.

\subsubsection{H2: Explanation preferences will vary between different redaction categories}

Taken together, these results support our hypothesis. Participants relied more on explanations and found them more helpful, when the system performed more redaction. When the system intervened heavily, explanations provided crucial context that aided participants in evaluating whether the output was trustworthy and appropriate. In contrast, when the system performed minimal modification, explanations were less necessary, offering little additional benefit. This pattern indicates that the usefulness of explanations is sensitive to the degree of system intervention and that provided explanations should scale with the amount of redaction performed.

These findings align with prior research showing that effective explanations must be tailored to context \cite{lim_why_2009}. Too little explanation can leave users uncertain about why a system behaved as it did, while too much can overwhelm or feel redundant when system behavior is already clear. Our results extend this principle to privacy-preserving systems, demonstrating that the optimal level of explanation detail shifts based on how substantially the system modifies content. This suggests the existence of a ``sweet spot'' in which explanations provide just enough insight to support user understanding without creating unnecessary cognitive load. More broadly, this pattern is consistent with work on progressive or selective transparency, which proposes that explanation detail should be revealed gradually depending on user needs and the system's level of intervention \cite{muralidhar_operationalizing_2025}. In our setting, explanations were particularly valued when redaction substantially changed the content, suggesting that explanation mechanisms for privacy-preserving systems may benefit from adapting based on the amount of modification performed by the system.

\subsection{Effect of Background and Experiences}
We finally analyze the results of the final part of the survey (Section~\ref{subsec:survey3}) to determine if there were any correlations between the background/experiences of the participants and their trust in the system. We used an aggregated trust score (averaged across the questions in Section \ref{subsec:survey1}) and grouped participants by explanation condition (no explanation, general explanation, and thorough explanation). Table \ref{tab:demographic_questions} shows the F-values, effect sizes, and p-values for each explanation style and for each demographic question.

\begin{table*}
\centering
\caption{Significant Results from the Background and Experience Survey Questions. Significant differences are indicated with one, two, and three asterices for $p<0.05$, $p<0.01$, and $p<0.001$, respectively.}
\label{tab:demographic_questions}
\resizebox{2\columnwidth}{!}{%
\begin{tblr}{
  cells = {c},
  cell{1}{2} = {c=3}{},
  cell{1}{5} = {c=3}{},
  cell{1}{8} = {c=3}{},
  vline{2-3,6} = {1}{},
  vline{2,5,8} = {2-14}{},
  hline{1,15} = {-}{0.08em},
  hline{3} = {-}{},
}
 & \textbf{No Explanation } &  &  & \textbf{General Explanation} &  &  & \textbf{Thorough Explanation} &  & \\
 & \textbf{F-value} & \textbf{Cohen's f} & \textbf{p-value} & \textbf{F-value} & \textbf{Cohen's f} & \textbf{p-value} & \textbf{F-value} & \textbf{Cohen's f} & \textbf{p-value}\\
\textbf{Age Group} &  &  &  &  &  &  & $F(5,52) = 3.12$ & $0.548$ & *\\
\textbf{Education} &  &  &  &  &  &  &  &  & \\
\textbf{Gender} &  &  &  &  &  &  &  &  & \\
{\textbf{Familiar with }\\\textbf{modern technology}} & $F(3,55) = 3.20$ & $0.416$ & * &  &  &  &  &  & \\
\textbf{Trust Technology} & $F(4,54) = 5.08$ & $0.613$ & * &  &  &  &  &  & \\
\textbf{Trust AI/LLM} & $F(3,55) = 3.55$ & $0.440$ & ** & $F(3,54) = 3.10$ & $0.415$ & * &  &  & \\
\textbf{Early technology adopter} & $F(3,55)=5.19$ & $0.532$ & ** &  &  &  &  &  & \\
{\textbf{Confident in learning }\\\textbf{new technologies}} & $F(3,55)=4.11$ & $0.474$ & * &  &  &  &  &  & \\
{\textbf{Trust output of }\\\textbf{system that they }\\\textbf{do not understand}} & $F(4,54) = 10.33$ & $0.875$ & *** & $F(4,53)=5.05$ & $0.618$ & ** &  &  & \\
{\textbf{Importance of }\\\textbf{understanding how}\\\textbf{~a system works}} &  &  &  &  &  &  &  &  & \\
{\textbf{Rely on AI system }\\\textbf{without understanding }\\\textbf{its reasoning}} &  &  &  & $F(4,53)=4.48$ & $0.582$ & ** &  &  & \\
{\textbf{Changed decisions }\\\textbf{based on AI output}} & $F(4,54)=3.49$ & $0.508$ & * &  &  &  &  &  & 
\end{tblr}
}
\end{table*}

\subsubsection{H3: People will have different levels of trust in the system based on their background and experiences}

In general, these findings provide partial support for this hypothesis. Some individual difference factors, e.g., baseline trust in technology and AI, were strongly predictive of trust. These effects were consistent across multiple measures and conditions, suggesting that participants enter AI-mediated tasks with predefined dispositions that shape their evaluations.

However, many other background variables did not produce significant effects, and the most significant findings were isolated to the no explanation condition. This pattern suggests that explanations play a stabilizing role on user trust, reducing the degree to which individual differences drive trust. When explanations were provided (general or thorough), trust levels became more uniform across demographic groups. Therefore, while certain personal dispositions matter, demographic characteristics alone (without considering explanations) are not predictors of trust, and their influence can be greatly minimized by an effective explanation design.

Our findings align with prior research showing that user characteristics, such as familiarity with technology, trust in AI systems, and comfort with automation, shape how individuals respond to AI-driven tools \cite{ehsan_who_2024}. In our study, trust was consistently higher among participants already predisposed to accept or rely on automation, regardless of the explanation condition. This pattern highlights a dual challenge for designers of privacy-preserving systems:

\begin{itemize}
    \item How do we maintain and appropriately calibrate the trust of users with a high baseline of confidence in AI without overwhelming them with unnecessary explanation?
    \item How do we build trust among users with lower baseline confidence, who may need more support or tailored explanations to feel comfortable with the system?
\end{itemize}

These findings suggest that adaptive or personalized explanation strategies may be essential. Systems that adjust explanation detail to user characteristics and provide accessible and contextually relevant information may better serve the goal of supporting trust across diverse user profiles.

%% file: text/06_conclusion.tex
As AI systems increasingly act as intermediaries in human communication, designing mechanisms that preserve privacy while maintaining user trust has become critical. This work examined how explanations can support trust in AI-mediated communication, particularly when sensitive information is redacted from messages. Our contributions included system development, technical validation, and user evaluation to explore the interplay between redaction, explanations, and individual differences in trust.

We demonstrated that providing explanations significantly improves participants' trust in the system's ability to preserve privacy. Participants who received explanations also engaged more and viewed their experience more positively with the system compared to those who did not receive explanations. Additionally, our results indicate that the effectiveness of explanations is context-dependent. Explanations are most valuable when the system performs extensive redactions, helping users understand the scope of the modifications and the reasoning behind them. When fewer modifications are necessary, the explanations are less critical, suggesting that optimal explanation strategies should adapt to the degree of system intervention.

Individual differences further affect trust in AI-mediated communication. Participants with more familiarity and trust in AI a priori were more likely to trust our system, but these differences were much more pronounced when explanations were not provided. Explanations were a stabilizing influence on user trust; trust levels became more uniform across demographic groups when explanations were provided. These findings highlight the importance of personalized or adaptive explanation strategies, which can tailor the level of detail to user characteristics and maximize trust without overwhelming or under-informing users.

In general, our results provide actionable insights for designing privacy-preserving AI intermediaries. Effective explanations strike a balance between transparency and privacy, offering enough detail to build trust while avoiding unnecessary exposure of sensitive information.

Future work should explore explanations in more diverse settings, including for different domains and more complex privacy protection systems, examine long-term trust evolution, and investigate adaptive explanation strategies that dynamically adjust to user characteristics and contextual information. Real-world trust dynamics may differ from the interactions explored in this work, particularly when users depend on redacted information to complete high-stakes tasks. Bridging this gap will require studying AI-mediated communication in more realistic environments, such as task-based settings where users must act on partially redacted information with real data constraints. Incorporating interactive and longitudinal evaluations can further capture how users adapt to redactions and explanations over time. By addressing these challenges, AI mediators can be designed to support privacy-preserving communication using explanations that improve user trust and system understanding.

Alternative approaches could include providing a standardized explanation along with labels indicating the categories of PII removed, or personalizing explanations by incorporating individualized information such as perceived threat severity or worst-case privacy scenarios. While these strategies may increase transparency, our findings suggest that explanation effectiveness strongly depends on contextual factors and individual differences. This highlights the value of developing adaptive explanation approaches that can select or tailor explanation styles based on the user and the context. Such mechanisms could enable AI mediators to support privacy-preserving communication while maintaining user trust and understanding.

\subsection{Limitations}
The user study focused on a controlled online setting with simulated communication scenarios, which may not fully capture the complexity of real-world interactions. We also collected the background and experience of the participants at the end of the study to avoid these questions affecting the responses of the participants during the study, but their background/experience responses were then affected by their study experience.

%% file: text/07_appendix.tex
\section{Appendix}
\subsection{Excerpt of example interaction between the \textcolor{CoordinatorColor}{\textbf{coordinator}} and \textcolor{CollaboratorColor}{\textbf{collaborator}}}
\label{sec:appendix-example_interaction}

\begin{figure}[ht]
    \centering
    \includegraphics[width=\columnwidth]{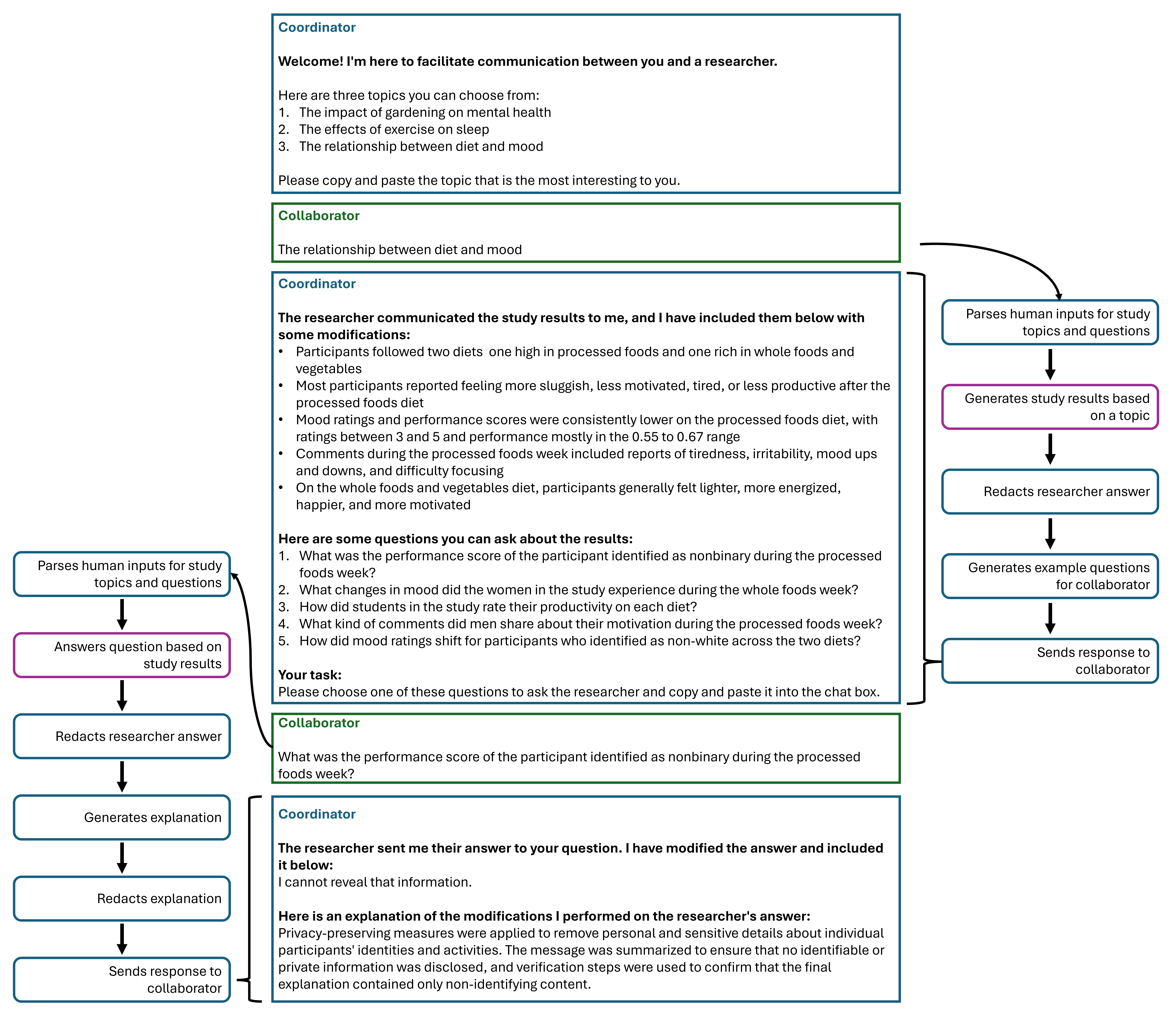}
    \caption{Excerpt of example interaction between the \textcolor{CoordinatorColor}{\textbf{coordinator}} and \textcolor{CollaboratorColor}{\textbf{collaborator}}. The conversation as well as tasks performed by the \textcolor{CoordinatorColor}{\textbf{coordinator}} and \textcolor{ResearcherColor}{\textbf{researcher}} in the system are indicated.}
    \label{fig:chat_outline}
\end{figure}

The interaction (Figure \ref{fig:chat_outline}) begins with the \textcolor{CoordinatorColor}{\textbf{coordinator}} showing a list of topics that the \textcolor{CollaboratorColor}{\textbf{collaborator}} can choose from, based on their interests. The \textcolor{CollaboratorColor}{\textbf{collaborator}} then chooses a topic, which in this case is the relationship between diet and mood. This begins a series of steps, shown to the right of the figure.
\begin{itemize}
    \item The \textcolor{CoordinatorColor}{\textbf{coordinator}} parses human input and passes the chosen study topic to the \textcolor{ResearcherColor}{\textbf{researcher}}.
    \item The \textcolor{ResearcherColor}{\textbf{researcher}} then generates study results based on that topic. These results include demographic information from the participants (e.g., age, gender, ethnicity), as well as data on their experiences during the experiment on mood with different diets.
    \item The \textcolor{CoordinatorColor}{\textbf{coordinator}} then removes private information from the study results and summarizes them for easy display to the \textcolor{CollaboratorColor}{\textbf{collaborator}}.
    \item The \textcolor{CoordinatorColor}{\textbf{coordinator}} generates a set of questions that could be interesting for the \textcolor{CollaboratorColor}{\textbf{collaborator}} to ask to get more information about the results.
    \item The \textcolor{CoordinatorColor}{\textbf{coordinator}} lastly compiles the summarized and redacted results and sample questions and sends them to the \textcolor{CollaboratorColor}{\textbf{collaborator}}.
\end{itemize}

The human \textcolor{CollaboratorColor}{\textbf{collaborator}} can then choose a question from the set of sample questions based on their interest, and this again begins a series of tasks taken by the \textcolor{ResearcherColor}{\textbf{researcher}} and \textcolor{CollaboratorColor}{\textbf{collaborator}}, shown to the left of the figure.
\begin{itemize}
    \item The \textcolor{CoordinatorColor}{\textbf{coordinator}} sends the question from the \textcolor{CollaboratorColor}{\textbf{collaborator}} to the \textcolor{ResearcherColor}{\textbf{researcher}}, who answers it based on the original study results.
    \item The \textcolor{CoordinatorColor}{\textbf{coordinator}} then redacts the \textcolor{ResearcherColor}{\textbf{researcher}}'s answer, removing any private information.
    \item The \textcolor{CoordinatorColor}{\textbf{coordinator}} generates an explanation of any operations that were performed on the original answer and removes any private information from the generated explanation.
    \item Lastly, the \textcolor{CoordinatorColor}{\textbf{coordinator}} compiles the redacted answer and explanation to send to the \textcolor{CollaboratorColor}{\textbf{collaborator}}.
\end{itemize}

\subsection{Prompting Details}
\label{sec:appendix-prompting_details}
We utilized GPT 4.1 (state-of-the-art at the time of the research) to generate study topics, original information, questions, and answers. We also utilized GPT 4.1 to redact sensitive information from both the original information and the answers. Due to regulations from our institution, we cannot include the full prompts for all generations, but for redaction, we instructed the LLM to remove all information that could potentially reveal the identity of an individual participant. For explanations, the LLM was instructed to generate an explanation of the specific style (general or thorough) that explains what the system did to transform the original information to the redacted information. We also utilized GPT 4.1 (with no history preserved) to validate redaction categories and redaction success for answers/original information (after LLM-judge was verified with human coders)

\subsection{Validating Redaction Categories}
\label{sec:appendix-redaction_categories}

\begin{figure}[t]
    \centering
    \includegraphics[width=\columnwidth]{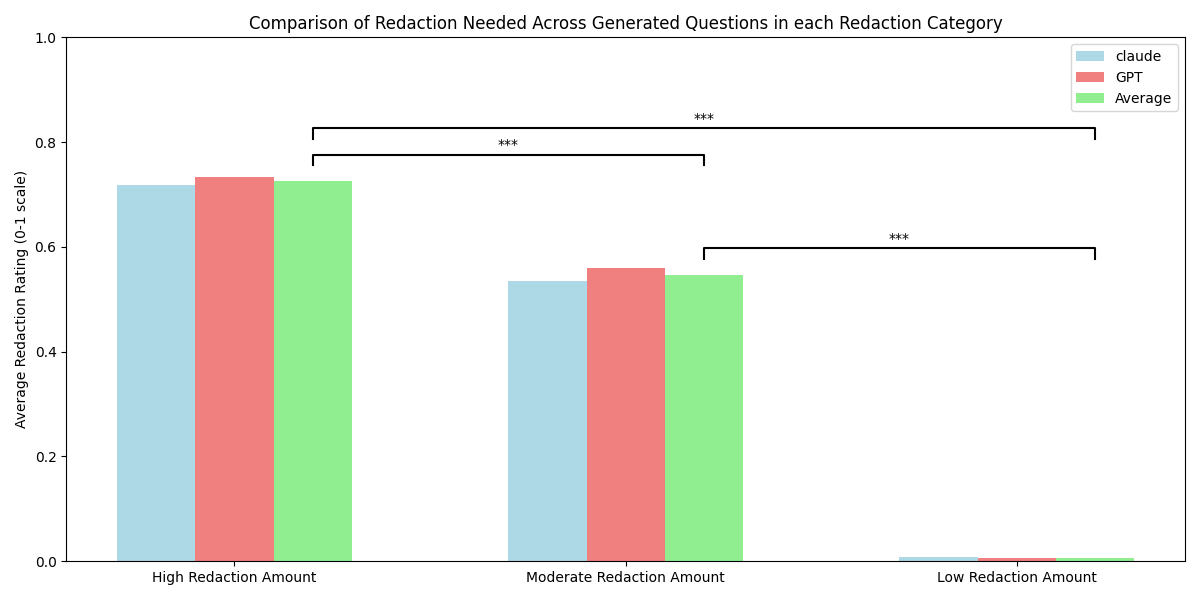}
    \caption{\textcolor{RevisionColor}{LLM judge (GPT 4.1, Claude Opus 4.7, and the average of the two) evaluation of amount of sensitive information across answers to questions in the three redaction categories: high, moderate, and low ($1$ indicates a large amount of sensitive information, $0$ indicates very little). Significant differences are shown with three asterices indicating $p<0.001$.}}
    \label{fig:redaction_amounts_violin_plot}
\end{figure}

We asked LLM judges (GPT 4.1 and Claude Opus 4.7) to rate the amount of sensitive information from $0$ to $1$ present in each researcher to questions in the high, moderate, and low redaction categories. \textcolor{RevisionColor}{The average ratings from each model for each redaction category are shown in Figure \ref{fig:redaction_amounts_violin_plot}, as well as the average of the two models. We can see the models are generally aligned in their judgment of the amount of redaction needed.}

We also performed an ANOVA on the average values and found that the categories require a significantly different amount of redaction ($F(2, 372) = 248.28, p<0.001$) A Tukey HSD post-hoc test resulted in all pairwise difference being significant at the $p<0.001$ level.

\subsection{Validating Privacy Protection of the Redacted Answer}
\label{sec:appendix-redacted_answer_privacy}

\begin{figure}[t]
    \centering
    \includegraphics[width=\columnwidth]{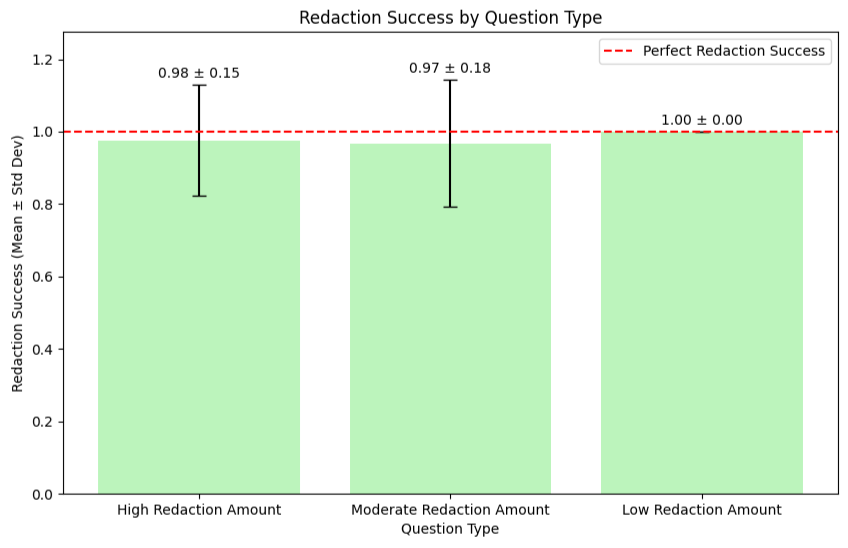}
    \caption{\textcolor{RevisionColor}{LLM judge (GPT 4.1) evaluation} of the privacy preservation success of our system to redact answers in the three redaction categories: high, moderate, and low ($1$ indicates privacy is protected, $0$ indicates privacy is not protected).}
    \label{fig:redaction_success_bar_plot}
\end{figure}

\textcolor{RevisionColor}{We asked a GPT4.1 LLM judge to determine whether privacy was preserved ($1$) or not ($0$) for each of the redacted answers. We previously validated that the LLM judge has high agreement with human coders and with another model (Claude Opus 4.7) for this task of detecting whether private information is being leaked.} Figure \ref{fig:redaction_success_bar_plot} shows the results of the LLM judge results, separated by category. We can see that our system performs extremely well in protecting privacy in the redacted answer.

\subsection{Explanation Examples}
\label{sec:appendix-explanation_examples}
Examples of these two explanation styles for the same answer and redacted answer are included below.

Example 1:
\begin{itemize}
    \item \textbf{General}: Identifying information, such as participant names, was removed from the original message to protect individual privacy. The message was then summarized to highlight the general experiences of participants sharing similar backgrounds, ensuring that all personally identifiable details and distinctive traits were excluded. The main findings were maintained in the summary, and the steps taken to ensure confidentiality were clearly outlined.
    \item \textbf{Thorough}: To preserve privacy, the original message was modified by removing direct identifiers and specific contextual details that could reveal individual identities. All personal names and characteristics were omitted, ensuring participants' confidentiality. Experiences were attributed to a collective group background rather than to individuals, maintaining the accuracy of findings without distorting key insights. Unnecessary individual-specific details were summarized or excluded, with core sentiments and comparative observations about gardening versus indoor activities retained. The process involved identifying and removing sensitive information, followed by checks to confirm that the essential meaning and clarity of the statement were preserved. The final message presents the collective experiences in compliance with privacy standards, upholding both confidentiality and clarity.
\end{itemize}

Example 2:
\begin{itemize}
    \item \textbf{General}: The participant's name and ethnicity were removed for privacy preservation and verification, and the message was summarized by omitting specific details while retaining the key information about sleep experiences before and after exercise.
    \item \textbf{Thorough}: In the privacy preservation process, personally identifiable information such as the participant's name and their specific regional identity were removed to protect their anonymity, ensuring the message cannot be traced back to an individual. Verification involved checking that all sensitive details had been appropriately omitted, while still maintaining the accuracy and context of the account described. Summarization further streamlined the message, retaining key points about the participant's sleep experiences with and without exercise, but excluding extraneous information irrelevant to the core findings. The overall process flow began with identifying sensitive personal data, then carefully editing the text to eliminate those elements, followed by reviewing the edited message for completeness and narrative integrity. As a result, the final message is both privacy-preserving and concise, allowing the participant's experiences to be shared safely and effectively within a broader context.
\end{itemize}

\subsection{Validating Privacy Protection of the Redacted Explanation}
\label{sec:appendix-redacted_explanation_privacy}

\begin{figure}[ht]
    \centering
    \includegraphics[width=\columnwidth]{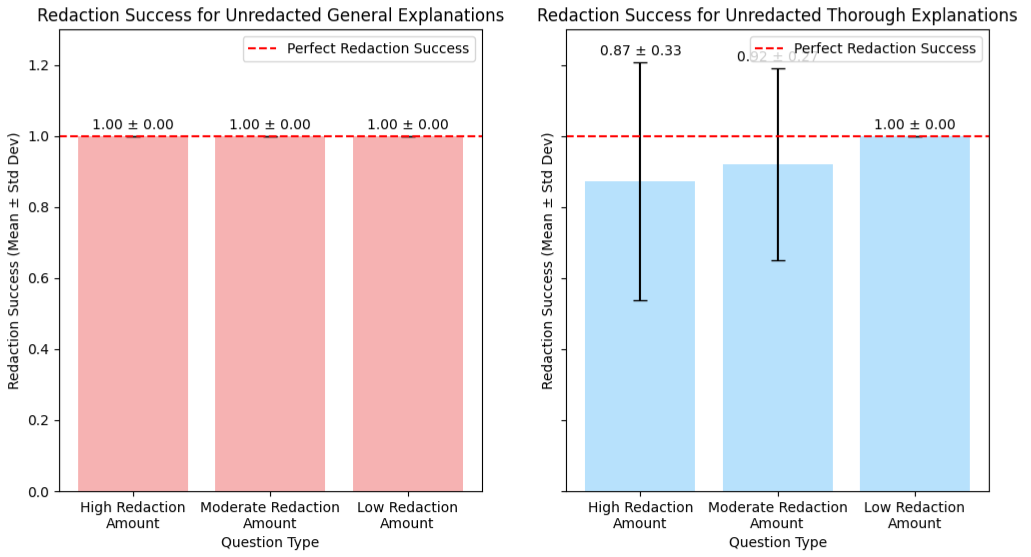}
    \caption{LLM judge evaluation of the privacy preservation success of our system to redact general and thorough explanations in the three redaction categories: high, moderate, and low ($1$ indicates privacy is protected, $0$ indicates privacy is not protected).}
    \label{fig:unredacted_explanation_redaction_success_bar_plot}
\end{figure}

\begin{figure}[ht]
    \centering
    \includegraphics[width=\columnwidth]{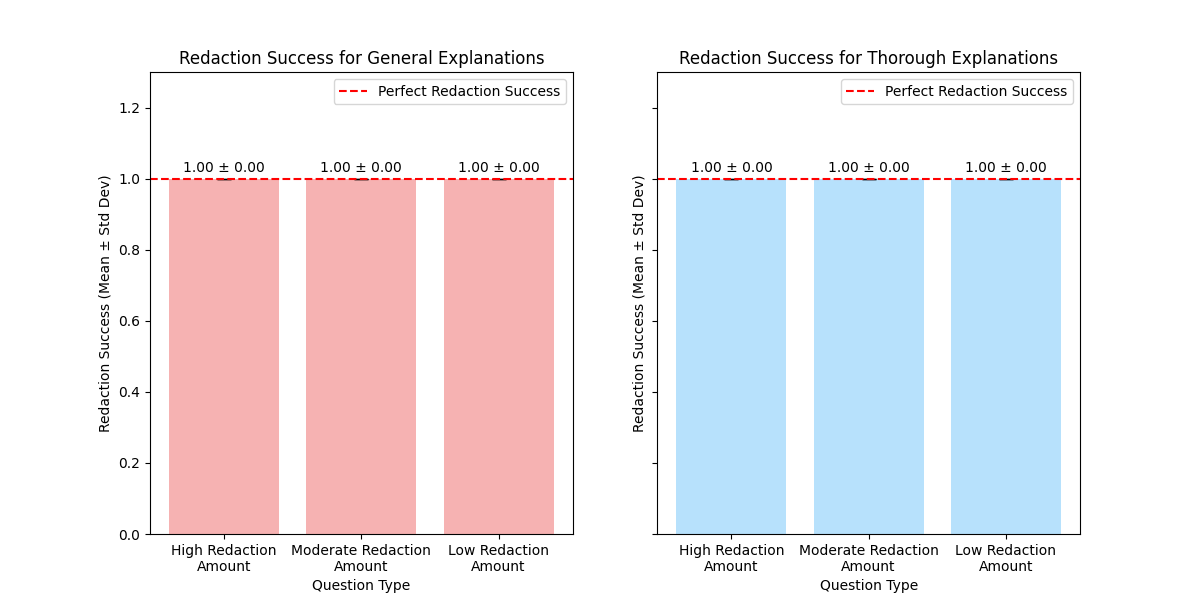}
    \caption{LLM judge evaluation of the privacy preservation success of our system to redact general and thorough explanations in the three redaction categories: high, moderate, and low ($1$ indicates privacy is protected, $0$ indicates privacy is not protected).}
    \label{fig:explanation_redaction_success_bar_plot}
\end{figure}

\textcolor{RevisionColor}{We asked a GPT 4.1 LLM judge to determine whether privacy was preserved ($1$) or not ($0$) for each of the unredacted and redacted explanations. We previously validated that the LLM judge has high agreement with human coders and with another model (Claude Opus 4.7) for this task of detecting whether private information is being leaked.} Figure \ref{fig:unredacted_explanation_redaction_success_bar_plot} shows the result of the LLM judge results, separated by category, of the private information in the unredacted explanations. We can see that there is no information leakage for general explanations. However, for thorough explanations, information leakage is occurring (more for the high redaction category and less for the low redaction category).

Figure \ref{fig:explanation_redaction_success_bar_plot} shows the results of the LLM judge results on the redacted explanations, separated by category. We can see that our system performs extremely well in protecting privacy in the redacted explanations.

\subsection{Survey Questionnaires}
\label{sec:appendix-questionnaire}

\subsubsection{Questions intended to explore explanation detail preferences}
\begin{enumerate}
    \item The researcher answer modified by the coordinator matched my expectations for a response to the question from you (strongly disagree to strongly agree)
    \item The explanation given by the coordinator (if provided) helped me understand how the system works (strongly disagree to strongly agree, N/A)
    \item The explanation given by the coordinator (if provided) had the appropriate amount of detail and was the appropriate length (far too little detail / much too short to far too much detail / much too long, N/A)
    \item The explanation given by the coordinator (if provided) helped me assess how accurate, reliable, or trustworthy the system is (strongly disagree to strongly agree, N/A)
    \item Please provide any additional comments or indicate if you wanted the system to explain anything else.
\end{enumerate}

\subsubsection{Questions intended to explore explanation detail and redaction amount interaction}
\begin{enumerate}
    \item I trust that the system preserved privacy appropriately over my entire interaction (strongly disagree to strongly agree)
    \item The explanations (if provided) over the entire interaction helped me understand what the system was doing (strongly disagree to strongly agree, N/A)
    \item I am confident in the system. I feel it works well (strongly disagree to strongly agree)
    \item The outputs of the system are very predictable (strongly disagree to strongly agree)
    \item The system is very reliable. I can count on it to be correct (strongly disagree to strongly agree)
    \item I feel safe that when I rely on the system, I will get the right answers (strongly disagree to strongly agree)
    \item The system can perform the redaction of private information better than a novice human user (strongly disagree to strongly agree)
    \item Please provide any additional comments or suggestions.
\end{enumerate}

\subsubsection{Questions intended to explore interaction between user characteristics and trust in our system}
\begin{enumerate}
    \item What is your age group?
    \item What is your highest level of education?
    \item What is your gender?
    \item How familiar are you with modern technology? (not at all familiar to extremely familiar)
    \item How much do you trust technology in general? (do not trust at all to completely trust)
    \item How much do you trust Artificial Intelligence (AI) or Large Language Models (LLM) systems? (do not trust at all to completely trust)
    \item How often do you use new technologies or digital tools before most people you know? (never to always)
    \item How confident are you in your ability to learn and use new technologies? (not at all confident to extremely confident)
    \item How much do you trust the output of a computer system when you do not understand how it works? (do not trust at all to completely trust)
    \item How important is it for you to understand how a system works before you trust its output? (not at all important to extremely important)
    \item How likely are you to rely on the output of an AI system if you do not understand its reasoning? (very unlikely to very likely)
    \item Have you ever changed your decision based on the output of an AI or automated system? (never to always)
\end{enumerate}